\documentclass[pra,twocolumn,superscriptaddress,amssymb]{revtex4}

\usepackage[colorlinks=true,linkcolor=magenta,citecolor=magenta,urlcolor=magenta,linktocpage=true]{hyperref}
\usepackage[utf8]{inputenc}
\usepackage[english]{babel}
\usepackage[T1]{fontenc}
\usepackage{amsmath}
\usepackage{cleveref}
\usepackage{svg}
\usepackage{tabularx}
\usepackage{array, makecell}
\usepackage[normalem]{ulem}

\usepackage{listings}
\usepackage{xcolor}
\definecolor{codegreen}{rgb}{0,0.6,0}
\definecolor{codegray}{rgb}{0.5,0.5,0.5}
\definecolor{codepurple}{rgb}{0.58,0,0.82}
\definecolor{backcolour}{rgb}{0.95,0.95,0.92}
\lstdefinestyle{mystyle}{
    backgroundcolor=\color{backcolour},   
    commentstyle=\color{codegreen},
    keywordstyle=\color{magenta},
    numberstyle=\tiny\color{codegray},
    stringstyle=\color{codepurple},
    basicstyle=\ttfamily\footnotesize,
    breakatwhitespace=false,         
    breaklines=true,                 
    captionpos=b,                    
    keepspaces=true,                 
    numbers=none,                    
    numbersep=5pt,                  
    showspaces=false,                
    showstringspaces=false,
    showtabs=false,                  
    tabsize=2
}
\usepackage[utf8]{inputenc}
\usepackage{newunicodechar}
\newunicodechar{，}{,}

\usepackage{psfrag,graphicx}
\usepackage{dcolumn}
\usepackage{bm}
\usepackage{amsfonts,amssymb,amsmath}        
\usepackage{slashed}
\usepackage[utf8]{inputenc}
\usepackage{graphicx}
\usepackage[caption=false]{subfig}
\usepackage{cancel}
\usepackage{xcolor}
\usepackage{amsmath}
\setcitestyle{numbers,square}

\newcommand{\be}{\begin{equation}}
\newcommand{\ee}{\end{equation}}
\newcommand{\bq}{\begin{eqnarray}}
\newcommand{\eq}{\end{eqnarray}}

\makeatletter

\renewenvironment{widetext@grid}{
  \par\ignorespaces
  \setbox\widetext@top\vbox{
   \vskip15\p@
   \hb@xt@\hsize{
    \leaders\hrule\hfil
    \vrule\@height6\p@
   }
   \vskip6\p@
  }
  \setbox\widetext@bot\hb@xt@\hsize{
    \vrule\@depth6\p@
    \leaders\hrule\hfil
  }
  \onecolumngrid

  \let\set@footnotewidth\set@footnotewidth@ii
}{
  \par

  \twocolumngrid\global\@ignoretrue
  \@endpetrue
}

\makeatother

\begin{document}
\title{Engineering non-Markovianity from spectral shaping of bosonic and fermionic baths}
\author{Si Luo}
\affiliation{State Key Laboratory of Porous Materials for Separation and Conversion \& MOE Key Laboratory of Computational Physical Sciences \& Research Center for Chemical Theory \& Department of Chemistry, Fudan University, Shanghai 200438, China}

\author{Cun Long}
\affiliation{Hefei National Research Center for Physical Sciences at the Microscale, University of Science and Technology of China, Hefei, Anhui 230026, China}

\author{Daochi Zhang}
\affiliation{State Key Laboratory of Porous Materials for Separation and Conversion \& MOE Key Laboratory of Computational Physical Sciences \& Research Center for Chemical Theory \& Department of Chemistry, Fudan University, Shanghai 200438, China}

\author{Massimiliano Di Ventra}
\email{diventra@physics.ucsd.edu}
\affiliation{Department of Physics, University of California San Diego, La Jolla, CA 92093}

\author{Xiao Zheng}
\email{xzheng@fudan.edu.cn}
\affiliation{State Key Laboratory of Porous Materials for Separation and Conversion \& MOE Key Laboratory of Computational Physical Sciences \& Research Center for Chemical Theory \& Department of Chemistry, Fudan University, Shanghai 200438, China}
\affiliation{Hefei National Laboratory, Hefei, Anhui 230088, China}

\date{\today}

\begin{abstract}
Non-Markovianity is the property that a physical system's future state depends on its previous dynamical history, namely the system has memory of its past. It is a key resource for quantum technologies but it is difficult to engineer without specific strategies. 
Here, we show that introducing a band gap in the environmental spectral density of both bosonic and fermionic baths provides a direct route to tuning non-Markovianity over a wide range. We demonstrate this effect by means of numerical
simulations across bosonic and fermionic environments and discuss experimental tests of our predictions. 



\end{abstract}
\maketitle
\vspace{5mm}
\newpage

\emph{Introduction.} Quantum systems inevitably interact with their environment~\cite{Petruccione,Gardiner,doi:10.1142/8334}. The latter plays a dual role: enabling essential operations like state preparation and sensing~\cite{RevModPhys.89.035002}, while inducing decoherence, a critical bottleneck for quantum computing and communication~\cite{Chuang,pittaluga_2025}. Depending on its spectral properties, the environment gives rise to either Markovian dynamics, where future evolution depends solely on the system's current state, or non-Markovian dynamics, characterized by memory effects that link the future to the system's past \cite{Rivas_2014,RevModPhys.88.021002,Xu26021002}. 


Non-Markovianity is not merely a dissipation byproduct but a pivotal quantum resource: it restores lost coherence~\cite{Panitchayangkoon2010,buluta_natural_2011,ishizakireview,PhysRevA.94.012110,wu_detecting_2020,PhysRevA.110.052220} and enhances entanglement~\cite{PhysRevLett.104.100502,PhysRevLett.104.250401,Ma_2012,PhysRevLett.108.160402,Friis_1,Chen2015,doi:10.1126/science.aax9743,Agusti2026}. This revival underpins advances from high-fidelity quantum transport~\cite{mui_enhanced_2025} and metrology beyond Markovian limits~\cite{doi:10.1126/science.1104149,PhysRevLett.127.060501,wfyl-wtz3} to quantum memristive elements for history-dependent switching~\cite{Zhe13086601,pfeiffer_quantum_2016,salmilehto_quantum_2017,spagnolo_experimental_2022,PhysRevApplied.18.024082,di2023memristors,Tan2023memristors}. Environmental memory also yields exotic phenomena such as the quantum Mpemba effect~\cite{PhysRevLett.134.220403}, accelerating state preparation~\cite{doi:10.1126/sciadv.adr4492,doi:10.1021/acs.jpclett.3c03159,PhysRevLett.134.050603}. Harnessing these effects is crucial for translating quantum advantages into practical devices.


Environmental memory effects have been extensively studied, with numerous metrics developed to quantitatively measure non-Markovianity~\cite{PhysRevLett.101.150402,PhysRevLett.103.210401,PhysRevA.82.042103,PhysRevLett.105.050403,PhysRevA.83.052128,Norambuena,PhysRevA.86.044101,PhysRevLett.116.020503,RevModPhys.89.041003,PhysRevA.97.012127}. 
These tools enabled systematic exploration of how thermodynamic and spectral parameters, such as temperature, bandwidth, and system-environment coupling strength, govern non-Markovianity~\cite{PhysRevLett.77.4728,PhysRevLett.88.197901}.  
While the dependence is often non‑monotonic~\cite{chen_using_2015,PhysRevA.102.022228,Wenderoth2021,Borrelli2014EffectOT}, strong non-Markovianity typically requires extreme conditions, such as ultra-low temperatures~\cite{PhysRevA.86.012115,RevModPhys.88.021002,PhysRevA.97.062104} or narrow bandwidths~\cite{PhysRevLett.108.160402,PhysRevA.86.012115,PhysRevA.102.022228}, posing significant challenges for practical implementations. Recent studies have increasingly focused on structured environments~\cite{PhysRevA.87.052328,PhysRevA.89.012114,PhysRevA.98.042125,kukita_controllable_2020,PhysRevLett.109.170402,hinarejos_non-markovianity_2017,PhysRevA.71.023812,PhysRevA.87.013428,Odeh2023nhg,w3vk-wx62}. 
For instance, spectral gaps in photonic and phononic crystals have been shown to robustly promote non-Markovian dynamics~\cite{PhysRevA.71.023812,PhysRevA.87.013428,Odeh2023nhg}. These works establish spectral engineering as a promising route to enhance non-Markovianity. However, despite these advances, a systematic strategy enabling continuous control of non-Markovianity remains lacking.

To resolve this challenge, here we demonstrate that non-Markovian
dynamics can be engineered via tailored spectral shaping of the environment spectral function. By introducing a sharp spectral edge at the system frequency, we establish a controllable mechanism to modulate non-Markovian dynamics through the edge steepness. We validate this strategy across both bosonic and fermionic environments, providing a unified framework for the continuous control of non-Markovianity. We finally discuss how to 
test our predictions experimentally.

\emph{Non-Markovianity measure.}
Consider a generic open quantum system coupled to a bath. The total Hamiltonian of the composite system reads $H_{\rm T} = H_{\rm S} + H_{\rm E} + H_{\rm I}$, where $H_{\rm S}$, $H_{\rm E}$, and $H_{\rm I}$ denote the Hamiltonians of the system, the bath environment, and the system-bath interaction, respectively. For a bosonic Gaussian bath with linear system-bath coupling $H_{\rm I} = -\hat{Q}\hat{F}$, where $\hat{Q}$ and $\hat{F}$ are Hermitian operators acting on the system and bath Hilbert spaces, respectively, the hybridization correlation function $C(t) \equiv \langle \hat{F}(t)\hat{F}(0)\rangle_{_{\rm E}} = {\rm tr}[\hat{F}(t) \hat{F}(0) \rho_{_{\rm E}}^{\rm eq}]$ fully encodes the temporal structure of environmental memory and dictates the resulting non-Markovian character of open-system dynamics~\cite{Yan05187,tanimura_reduced_2014,song_calculation_2015,zhang_imaginary-time_2022}. Here, $\hat{F}(t) = e^{iH_{\rm E}t} \hat{F} e^{-iH_{\rm E}t}$, $\rho_{_{\rm E}}^{\rm eq} = e^{-\beta H_{\rm E}}/{\rm tr}(e^{-\beta H_{\rm E}})$ is the equilibrium density matrix of the bare bath at inverse temperature $\beta = 1/T$, and we set $\hbar = k_{\rm B} = 1$ throughout. The correlation function $C(t)$ is entirely determined by the bath spectral function, $J(\omega) \equiv \frac{1}{2}\int_{-\infty}^\infty \langle [\hat{F}(t), \hat{F}(0)]\rangle_{_{\rm E}} e^{i\omega t} dt$, via the fluctuation-dissipation theorem \cite{Yan05187}:
\begin{equation}  \label{eq:correlation}
C(t)=\frac{1}{\pi}\int_{-\infty}^\infty  J(\omega)f_\beta(\omega) \,e^{-i\omega t}d\omega,
\end{equation}
where $f_\beta(\omega)=[1-\exp(-\beta\omega)]^{-1}$ is the Bose-Einstein distribution function. An analogous relation holds for fermionic baths, establishing the same direct mapping between the bath spectral function and environmental memory for fermionic baths.

At finite temperatures, the bath correlation function exhibits exponential decay, $C(t) \sim e^{-t/\tau_c}$, with a decay timescale dictated by the temperature, i.e., $\tau_c \propto \beta$~\cite{SM}.  
Consequently, at high temperatures $\tau_c$ can be considerably shorter than the characteristic timescale of the system dynamics, and the reduced dynamics is well described by a Markovian Lindblad master equation \cite{Lindblad}. In contrast, as $T \rightarrow 0$, $\tau_c$ diverges. For bath spectral functions that are smooth at the chemical potential, $C(t)$ exhibits algebraic decay in the zero-temperature limit, i.e., $C(t) \sim 1/t$ for a fermionic bath and $C(t) \sim 1/t^{1+\alpha}$ for a bosonic bath with $J(\omega) \propto \omega^{\alpha}$ in the low-energy regime \cite{Giraldi2017,Aha18104306,Bha22128226}.
This slow decay invalidates the Markovian approximation, giving rise to pronounced memory effects and non-Markovian dynamics.

To quantify non-Markovianity in open quantum dynamics governed by the reduced density matrix $\rho(t) = {\rm tr}_{\rm E}(\rho_{\rm T})$, where $\rho_{\rm T}$ is the total density matrix and ${\rm tr}_{\rm E}$ denotes the partial trace over the bath environment, we adopt the Breuer-Laine-Piilo (BLP) measure \cite{PhysRevLett.103.210401},
\begin{equation}\label{eq:nonM}
\mathcal{N}=\underset{\rho_{1,2}(0)}{\rm max}\underset{dD(t)/dt>0}{ \int} \frac{d D(t)}{d t} \,dt.    
\end{equation}
Here, $D(t)=\|\rho_1(t)-\rho_2(t)\|/2$ is the trace distance between states evolved from the initial pair $\rho_{1,2}(0)$, with $\|A\|={\rm tr}\sqrt{A^\dagger A}$ the trace norm. Physically, $\mathcal{N}$ quantifies the total extent of information backflow from the environment to the system; a vanishing $\mathcal{N}$ signifies Markovian evolution, characterized by a strictly monotonic decrease in $D(t)$ ($d D(t)/dt\leq0$). Although rigorous evaluation of $\mathcal{N}$ requires maximization over all initial-state pairs $\rho_{1,2}(0)$, previous work \cite{PhysRevA.86.062108} has establishes that the optimal pair invariably consists of orthogonal states residing on the boundary of the state space. Leveraging this property, we select appropriate initial states {\it a priori} in our numerical simulations, thereby circumventing computationally prohibitive optimization routines.

 \begin{figure}[t!]
    \centering
    \includegraphics[width=1\linewidth]{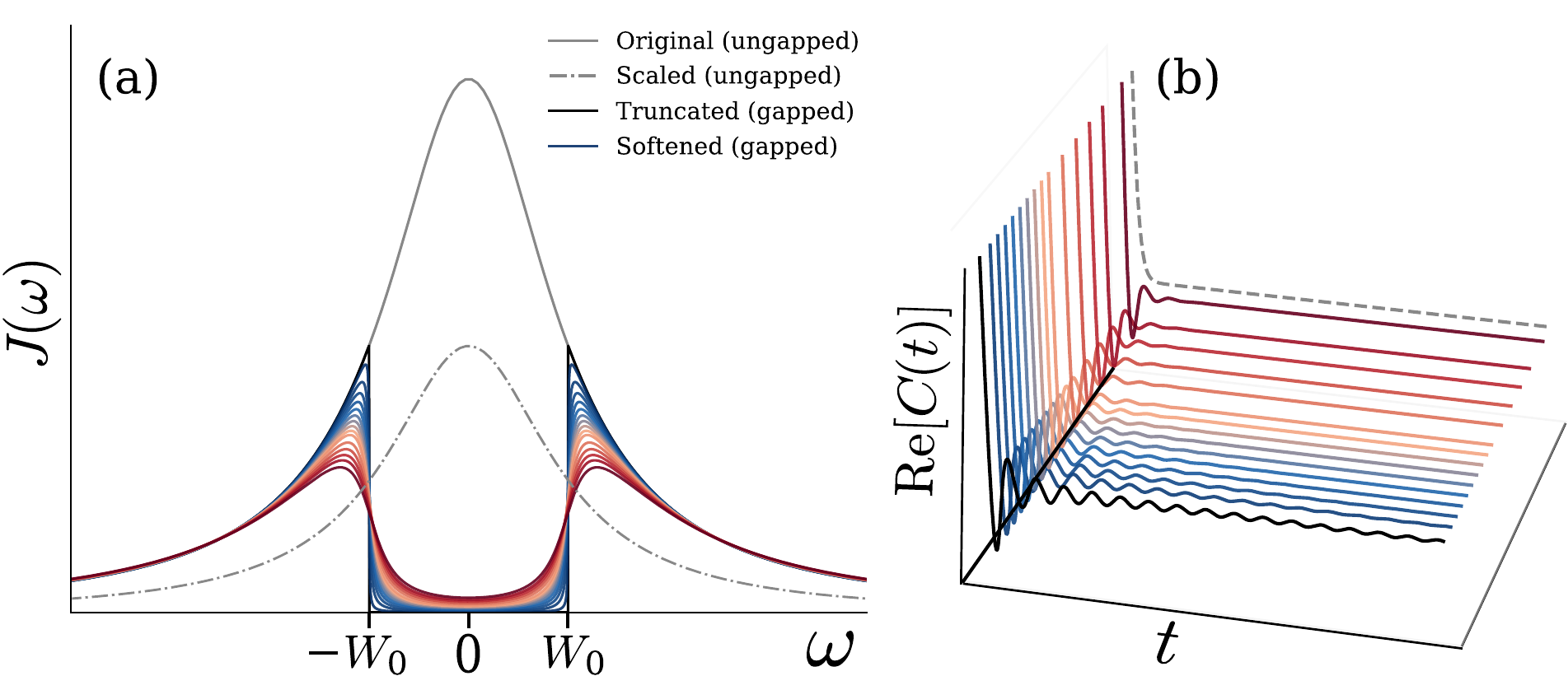}
    \caption{Spectrum truncation and correlation functions for a fermionic bath. (a) Original bath spectral function (gray), truncated spectral function with a band gap of width $2W_0$ (black), softened gapped spectral functions (colored), and a scaled ungapped spectral function (gray dash-dot). All spectral functions have the same area below the curves except the original one. (b) Real part of the corresponding environment correlation functions.}
    \label{fig:bandgapS}
\end{figure}

\emph{Strategy for enhancing non-Markovianity.}
To enhance non-Markovianity, we tailor the bath spectral function $J(\omega)$ to prolong the decay timescale of $C(t)$ via Eq.~\eqref{eq:correlation} \cite{Ma2014,Wang2017,lemma}. While conventional strategies, such as lowering temperature or narrowing bandwidth, extend the correlation time, they often inadvertently modify the integrated system-bath hybridization strength  $\frac{1}{\pi}\int_{-\infty}^\infty J(\omega)f_\beta(\omega)d\omega$. This alteration accounts for the non-monotonic variation of non-Markovianity with respect to the controlling parameter reported in prior studies \cite{chen_using_2015,PhysRevA.102.022228,Wenderoth2021,Borrelli2014EffectOT}. Furthermore, a prolonged correlation time alone is insufficient to ensure information backflow. For instance, zero-frequency modes retain information indefinitely yet preclude its return due to the absence of oscillatory components \cite{Addis2014,pure_dephasing}. Instead, backflow requires coherent oscillatory exchange of information between the system and environment, which necessitates an oscillatory lineshape of $C(t)$. Consequently, achieving strong non-Markovianity requires satisfying two distinct conditions: a sufficiently slow decay to retain information, and oscillations resonant with the system frequency to facilitate its coherent revival.

To satisfy these dual requirements, we introduce a gap at zero energy in the original bath spectral function $J_0(\omega)$, such that (see Fig.~\ref{fig:bandgapS} for the example of a fermionic bath)
\begin{equation} \label{eqn:def-jw}
 J_{\rm trun}(\omega)=\Theta(|\omega|-W_0)J_0(\omega).
\end{equation}
Here, the Heaviside step function $\Theta(x)$ abruptly truncates the spectral function at $\pm W_0$, effectively decoupling the system from the bath within this interval. Crucially, the steep edges of this gap induce an algebraic decay modulated by coherent oscillations, yielding the asymptotic form of the correlation function (see the Supplemental Material (SM) \cite{SM} for details):
\begin{equation} \label{eqn:ct-gap-1}
   C_{\rm trun}(t) \sim \frac{e^{-iW_0 t}}{t}.
\end{equation}
Here, while the $1/t$ decay persists irrespective of temperature, the oscillatory factor $e^{-iW_0t}$ dominates only at sufficiently low temperatures ($T\ll W_0$) \cite{SM}. Notably, Eq.~\eqref{eqn:ct-gap-1} holds for both bosonic and fermionic baths, because the zero-energy gap nullifies the influence of the distribution function $f_\beta(\omega)$
through which the difference in statistics manifests in $C(t)$ via Eq.~\eqref{eq:correlation}. The form of Eq.~\eqref{eqn:ct-gap-1} intrinsically fulfills both prerequisites for strong non-Markovianity: 
the $1/t$ envelop ensures slow information decay, while the $e^{-iW_0 t}$ phase enables resonant backflow.  Consequently, when the band-edge energy $W_0$ is tuned to match the system frequency $\omega_s$ critical to the dissipative dynamics, the information backflow is maximized \cite{PhysRevA.89.022109}.


In realistic systems, the band edge exhibits a finite yet tunable steepness ~\cite{YuCardona2005,Kittel2005,ionescu_tunnel_2011,chen_robust_2020,icking_ultrasteep_2024}, which allows for continuous control over the long-time decaying trend of $C(t)$. Theoretically this can be realized via an edge-softening protocol by convolving the truncated spectral function $J_{\rm trun}(\omega)$ with a smearing kernel $g(\omega)$ as follows, 
\begin{equation} \label{eq:Lorentz_moll}
J(\omega) = \int_{-\infty}^\infty J_{\rm trun}(\omega') g(\omega-\omega') \,d\omega'.
\end{equation}


As shown in Fig.~\ref{fig:bandgapS}, for a Lorentzian kernel $g(\omega)$ of width $\epsilon$ (hereafter referred to as the softening width), the correlation function exhibits a long-time oscillatory decay of
$C(t) \sim e^{-\epsilon t}e^{-iW_0 t}\,t^{-1}$ under the condition $T \ll W_0$. Such an asymptotic behavior holds for both bosonic and fermionic baths (see the SM \cite{SM}). 
Consequently, the decay timescale decreases monotonically with increasing $\epsilon$, corresponding to progressive smoothing of the band edge. Importantly, $\frac{1}{\pi}\int_{-\infty}^\infty J(\omega) f_\beta(\omega)d\omega$ remains invariant upon varying $\epsilon$. This decouples the correlation time from the total hybridization strength, ensuring that tuning the edge steepness modulates only the memory timescale and yields clean, monotonic control over non-Markovianity. Crucially, this spectral-shaping strategy is universally applicable to both bosonic and fermionic environments, as we demonstrate below.

\emph{Open quantum system models.} 
We illustrate our strategy using two paradigmatic models.
First, we consider a spin-boson model (SBM) comprising a two-level system described by $H_{\rm S} = \omega_s \hat{\sigma}_z / 2$, where $\hat{\sigma}_z$ is the Pauli-z matrix and $\omega_s$ the level splitting. The bosonic environment is modeled as a collection of harmonic oscillators, $H_{\rm E}=\sum_k\omega_k \hat{b}_k^\dagger \hat{b}_k$, with $\hat{b}_k^\dagger$ ($\hat{b}_k$) creating (annihilating) a mode of frequency $\omega_k$. The system–bath interaction is $H_{\rm I} = \hat{\sigma}_x \sum_k g_k (\hat{b}_k + \hat{b}_k^\dagger)$, where $\hat{\sigma}_x$ is the Pauli-x matrix and $g_k$ denotes the coupling strength.

Second, we examine a single-impurity Anderson model (SIAM), wherein a local orbital couples to a fermionic reservoir. The system Hamiltonian reads $H_{\rm S} = -\omega_s \sum_{\alpha=\uparrow,\downarrow} \hat{n}_\alpha + U \hat{n}_\uparrow \hat{n}_\downarrow$, with on-site energy $-\omega_s$, 
Coulomb repulsion $U = 2\omega_s$, and occupation number $\hat{n}_\alpha = \hat{d}_\alpha^\dagger \hat{d}_\alpha$; here, $\hat{d}_\alpha^\dagger$ ($\hat{d}_\alpha$) creates (annihilates) a spin-$\alpha$ electron on the local orbital. The reservoir is described by $H_{\rm E}=\sum_{k\alpha} \epsilon_{k\alpha} \hat{a}_{k\alpha}^\dagger \hat{a}_{k\alpha}$, where $\epsilon_{k\alpha}$ denotes the dispersion of the spin-$\alpha$ band and $\hat{a}_{k\alpha}^\dagger$  ($\hat{a}_{k\alpha}$) are the corresponding creation (annihilation) operators. The system-reservoir interaction is $H_{\rm I} = \sum_{k\alpha} g_{k} \hat{a}_{k\alpha}^\dagger \hat{d}_\alpha + {\rm H.c.}$, with $g_{k}$ being the coupling strength. 

In both models, the environments obey Gaussian statistics, so that the non-Markovianity of the reduced system dynamics is largely dictated by the bath spectral function $J(\omega)$ via Eq.~\eqref{eq:correlation}. 
We assume an underdamped Brownian form, $J_0(\omega) = \lambda\gamma\omega_0^2 \omega / [(\omega^2-\omega_0^2)^2+\gamma^2\omega^2]$, for the SBM, and a Lorentzian form, $J_0(\omega) = \lambda\gamma^2/(\omega^2+\gamma^2)$, for the SIAM. Here, $\lambda$ denotes the system-bath hybridization strength, $\gamma$ the bandwidth, and $\omega_0$ the characteristic frequency of the bosonic bath. The gapped spectral function $J(\omega)$ is then obtained by applying the band-gap truncation [Eq.~\eqref{eqn:def-jw}] followed by the edge-softening convolution [Eq.~\eqref{eq:Lorentz_moll}]. For simplicity, we adopt $\lambda=\gamma=\omega_0=W_0 = \omega_s$ throughout, and set the temperature to $T=\omega_s/1000$.

All open quantum dynamics simulations are performed using the numerically exact hierarchical equations of motion (HEOM) method \cite{Tanimura_3,Yan_5,tanimura_numerically_2020} implemented in MOSCAL \cite{MOSCAL10,MOSCAL20} and HEOM-QUICK \cite{Ye16608,quick}. The Prony spectral fitting algorithm \cite{chen_universal_2022} is employed for the accurate and efficient evaluation of decomposition of $C(t)$.

\begin{figure}[t!]
\centering
\includegraphics[width=\linewidth]{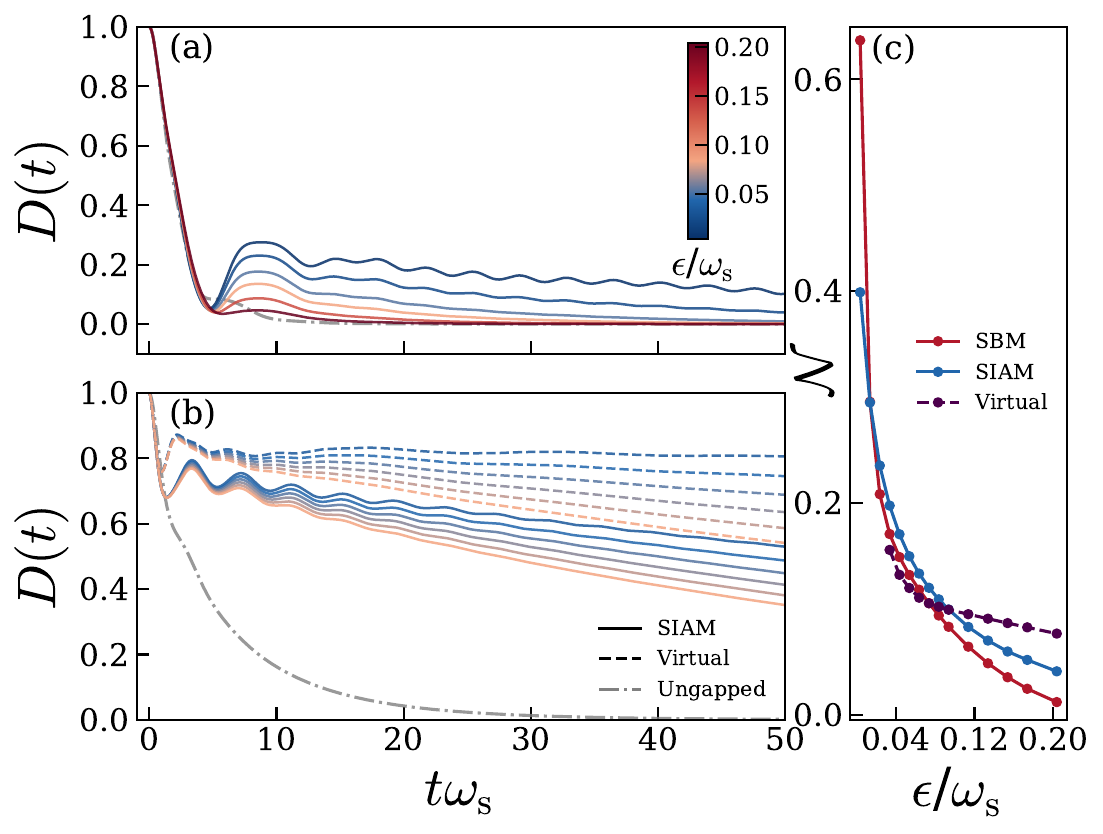}
\caption{Non-Markovianity control via environmental spectrum engineering. Trace distance dynamics $D(t)$ for (a) an SBM and (b) an SIAM at selected softening widths $\epsilon$ (color-coded). The parameter $\epsilon$ controls the band edge sharpness [cf. Eq.~\eqref{eq:Lorentz_moll}]. The gray dash-dotted lines in (a) and (b) provide a benchmark for a scaled Lorentzian (ungapped) bath spectral function. (c) Non-Markovianity measure $\mathcal{N}$ versus $\epsilon$ for both models. The purple curve shows results from a virtual SIAM simulation with the fermionic parity replaced by the bosonic parity within the HEOM method. Model parameters are given in the main text.}
\label{fig:nonm_eps}
\end{figure}

\emph{Non-Markovian effects on the short-time dynamics.}
At short times, we track the trace distance $D(t)$ between two initial system states: 
$|{\uparrow}\rangle$ and $|{\downarrow}\rangle$, which represent the spin-up and spin-down states for the SBM and the two singly occupied states for the SIAM. Increases in $D(t)$ indicate information backflow from the bath to the system, marking the emergence of non-Markovian effects. Figure~\ref{fig:nonm_eps}(a) and (b) show the results for both models. For reference, the benchmark of a scaled ungapped bath is also included, which yields a vanishing BLP measure ($\mathcal{N} = 0$). As the softening width $\epsilon$ increases, $D(t)$ decays smoothly over time, and the intervals exhibiting increases in $D(t)$ shrink progressively. Consequently, $\mathcal{N}$ decays monotonically with $\epsilon$ [Fig.~\ref{fig:nonm_eps}(c)], with the most dramatic drop occurring near $\epsilon=0$ as the band edge softens away from the truncated limit.

Despite this common trend, the two models differ in both the dynamical pattern of $D(t)$ and the sensitivity of non-Markovianity to $\epsilon$, reflecting their distinct model structures and quantum statistics. To isolate the role of statistics, we perform a virtual experiment by replacing the fermionic parity with the bosonic counterpart by discarding all $(-1)$ factors arising from Grassmann anticommutation in the fermionic HEOM \cite{Han18234108}. 
As shown in Fig.~\ref{fig:nonm_eps}(c), this substitution reveals that fermionic parity makes non-Markovianity markedly more sensitive to $\epsilon$: ignoring fermionic parity suppresses $\mathcal{N}$ at small $\epsilon$. Consistently, Fig.~\ref{fig:nonm_eps}(b) shows that the solid lines (physical fermionic case) exhibit more pronounced short-time oscillations than the dashed lines (virtual bosonic parity) for the same $\epsilon$, indicating stronger information backflow and thus stronger non-Markovianity.

We now generalize the single-spin SBM to a two-spin setup to explore the interplay between non-Markovianity and entanglement. The system, described by the Hamiltonian $H_{\rm S} = \omega_s(\hat{\sigma}_{z,1} + \hat{\sigma}_{z,2})/2$, couples to the same bath as above via the interaction $H_{\rm I} = (\hat{\sigma}_{x,1} + \hat{\sigma}_{x,2})\sum_k g_k(\hat{b}_k + \hat{b}_k^\dagger)$.
The system is initialized in a separable state $\left\lvert\uparrow\right\rangle_1 \otimes \left\lvert\downarrow\right\rangle_2$ and then allowed to evolve freely. We track the temporal evolution of the inter-spin entanglement by probing the concurrence $\mathcal{C}(\rho) \equiv \max(0, \sqrt{\lambda_1} - \sqrt{\lambda_2} - \sqrt{\lambda_3} - \sqrt{\lambda_4})$ \cite{Wootters1998}, where $\lambda_i$ are the eigenvalues of $R = \rho(\sigma_{y} \otimes \sigma_{y})\rho^\ast (\sigma_{y} \otimes \sigma_{y})$ in decreasing order.

\begin{figure}[t!]
    \centering
\includegraphics[width=1\linewidth]{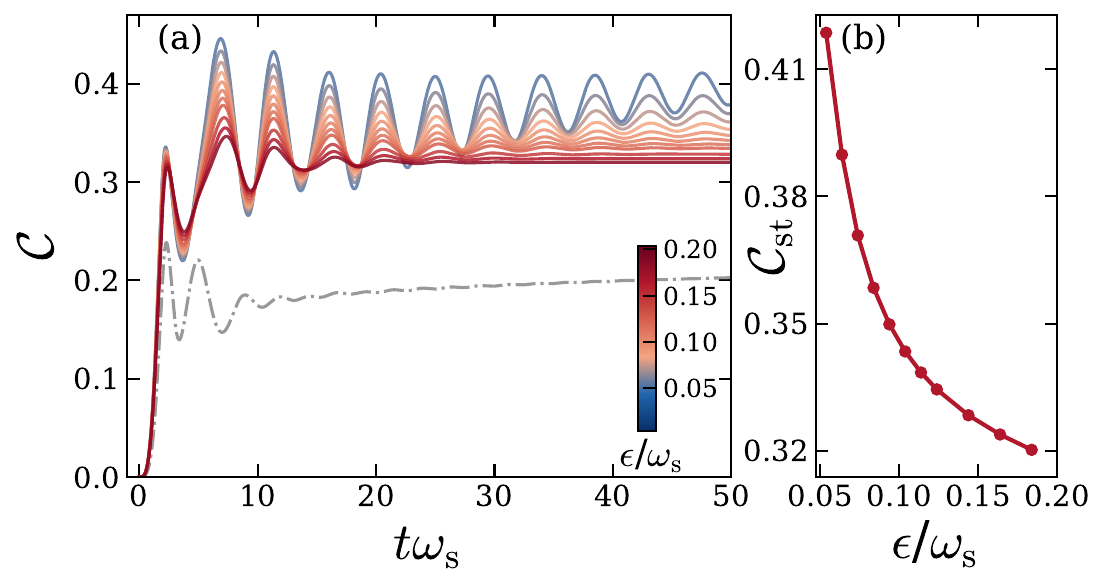}
\caption{Concurrence dynamics of two spins coupled to a common bath with tunable softening width $\epsilon$. (a) Time evolution for selected $\epsilon$ (colored) and the benchmark for a scaled ungapped bath (gray dash-dotted). (b) Steady-state concurrence $\mathcal{C}_{\rm st}$ as a function of $\epsilon/\omega_s$. Model parameters are given in the main text.}
    \label{fig:entanglement}
\end{figure}

Figure~\ref{fig:entanglement}(a) shows the concurrence dynamics for different softening widths $\epsilon$. The concurrence emerges from zero at short times and approaches a finite steady-state value for each $\epsilon$, in contrast to the benchmark for a scaled ungapped bath, which yields much weaker entanglement. Clearly, smaller $\epsilon$ leads to stronger concurrence and slower decay of transient oscillations. Figure~\ref{fig:entanglement}(b) reveals a monotonic increase of the steady-state concurrence as $\epsilon$ decreases, with the most pronounced enhancement occurring in the small-$\epsilon$ regime. Importantly, since there is no direct coupling between the two spins, the entanglement arises entirely from the couplings to a common bath and is thus strongly governed by the environmental memory content.

\emph{Non-Markovian effects on the long-time dynamics.}
Beyond the short-time entanglement revival, non-Markovianity also significantly affects the long-time dynamics. To illustrate this, we examine the spin polarization of the SBM and SIAM, defined as $P(t) = [n_\uparrow(t) - n_\downarrow(t)]/[n_\uparrow(t) + n_\downarrow(t)]$, where $n_{\uparrow(\downarrow)}(t)$ are the occupation numbers of the spin-up (down) states. The system is initialized in $\vert{\downarrow}\rangle$  for both models. The relaxation time $T_{\rm r}$ is extracted from the decay of $P(t)$ to its steady-state value $P_{\rm st}$ via 
$T_{\rm r}={\rm min}\left\{t \;\middle|\; \sup_{t' \geq t} \frac{|P(t')-P_{\rm st}|}{|P(0)-P_{\rm st}|} < 10^{-3}\right\}$.

\begin{figure}[t!]
\centering
\includegraphics[width=\linewidth]{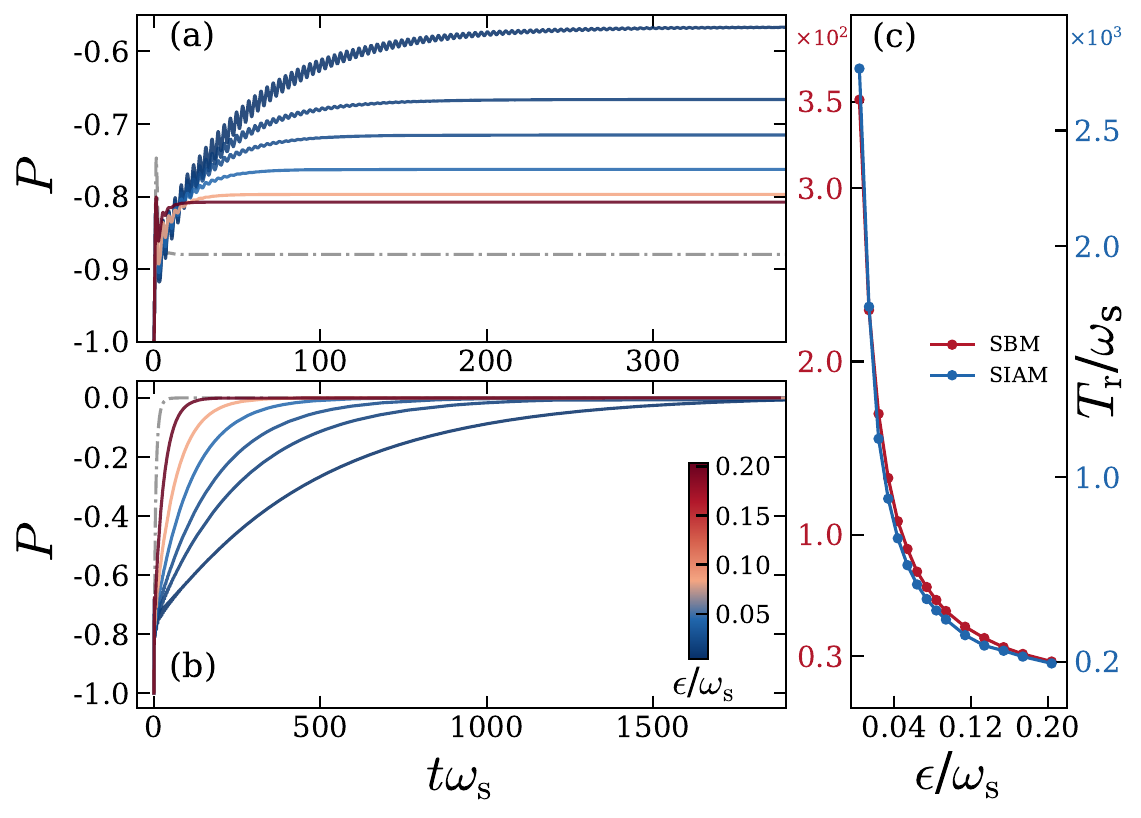}
\caption{Relaxation dynamics of spin polarization in (a) an SBM and (b) an SIAM for selected softening widths $\epsilon$. Gray dash-dotted lines are benchmarks for the scaled ungapped bath. (c) Relaxation time $T_{\rm r}$ versus $\epsilon$ for both models; the two curves exhibit a similar dependence on $\epsilon$ up to an overall scale factor. Model parameters are given in the main text.}
\label{fig:relaxT}
\end{figure}

We then investigate the relaxation dynamics. As shown in Fig.~\ref{fig:relaxT}(a) and (b), the introduction of a gap substantially extends the relaxation time in both models, whereas the ungapped bosonic benchmark (gray dash-dotted in Fig.~\ref{fig:relaxT}(a)) reaches its stationary value as early as \(t\omega_s \simeq 15\). Specifically, a smaller $\epsilon$ systematically slows down the spin-polarization relaxation. Moreover, as shown in Fig.~\ref{fig:relaxT}(c), the relaxation time $T_{\rm r}$  exhibits a similar nonlinear dependence on $\epsilon$ up to an overall scale factor, with the fermionic relaxation time systematically longer. This reflects the underlying quantum statistics: Pauli blocking suppresses spin redistribution, leading to intrinsically slower relaxation in the SIAM. These results confirm that engineering the bath spectral function effectively, monotonically, and substantially tunes the long-time quantum dissipative dynamics.

\emph{Experimental Feasibility.}
Our strategy for engineering non-Markovianity via spectral shaping can be potentially realized in experiments. The ideal truncated spectral density can be approximated by environments featuring a low density-of-states (DOS) window centered at the reference energy, the Fermi level for electrons or zero frequency for photons, whose boundaries are sharp van Hove singularities (VHS). Such DOS profiles can be achieved in moir\'e superlattices, periodic structures formed by stacking two crystal layers with a relative twist. The long-period moir\'e potential creates a low-DOS window flanked by VHS, with both the window width and VHS sharpness tunable \cite{Yuan2019, Yi2022, Bampoulis_2024}. For electrons, moir\'e superlattices have been realized in topological insulator thin films, where the surface-state DOS vanishes at the Fermi level \cite{Grasser2025, Zhang2025, Song2025}. The twist angle controls both the gap width and VHS sharpness \cite{Yuan2019, Zhang2019, Nakatsuji2023}, while electrostatic gating shifts the low-DOS window without altering its profile \cite{Xiong2012, Ireland2018, Wu2026}. For photons, photonic moir\'e superlattices offer an analogue, with the DOS vanishing at zero frequency and diverging near VHS \cite{Wang2023, Wang2025}, where the gap width and edge steepness are similarly tunable via the twist angle or dielectric modulation \cite{Yi2022, Wang2023}.

\emph{Conclusion.} To summarize, introducing an energy gap with tunable edge steepness in the bath spectral function provides direct, continuous control over non-Markovianity in both bosonic and fermionic baths. Importantly, since the gap is centered at zero energy, the reshaped spectral function remains effectively temperature-insensitive as long as $T \ll W_0$, enabling strong non-Markovianity without requiring ultra-low temperatures. Across the studied models, the dependence on edge smoothness is universal: sharper edges enhance non-Markovianity, and generates stronger bath-mediated entanglement and slower relaxation, with fermionic statistics imposing a longer timescale. Experimentally, this control is accessible in moir\'e superlattices, where the tunable DOS can be used to realize the engineered spectral density. We thus hope this work will motivate such experiments.

\begin{acknowledgments}

S.L., C.L, D.Z., and X.Z. are grateful for the funding support from the Quantum Science and Technology -- National Science and Technology
Major Project (Grant No. 
2021ZD0303306), and the National Natural Science Foundation of China (Grant Nos. 22425301, 22393912, and 22503019).  No U.C. San Diego or U.S. federal funding have been used in this collaboration. No U.C. San Diego resources or data were used in this work.

\end{acknowledgments}

\nocite{apsrev41Control}
\bibliographystyle{apsrev4-2}
\bibliography{bib}
\end{document}


\title{ Engineering non-Markovianity from spectral shaping of bosonic and fermionic baths}

\author{Si Luo}
\affiliation{State Key Laboratory of Porous Materials for Separation and Conversion \& MOE Key Laboratory of Computational Physical Sciences \& Research Center for Chemical Theory \& Department of Chemistry, Fudan University, Shanghai 200438, China}

\author{Cun Long}
\affiliation{Hefei National Research Center for Physical Sciences at the Microscale, University of Science and Technology of China, Hefei, Anhui 230026, China}

\author{Daochi Zhang}
\affiliation{State Key Laboratory of Porous Materials for Separation and Conversion \& MOE Key Laboratory of Computational Physical Sciences \& Research Center for Chemical Theory \& Department of Chemistry, Fudan University, Shanghai 200438, China}

\author{Massimiliano Di Ventra}
\email{diventra@physics.ucsd.edu}
\affiliation{Department of Physics, University of California San Diego, La Jolla, CA 92093}

\author{Xiao Zheng}
\email{xzheng@fudan.edu.cn}
\affiliation{State Key Laboratory of Porous Materials for Separation and Conversion \& MOE Key Laboratory of Computational Physical Sciences \& Research Center for Chemical Theory \& Department of Chemistry, Fudan University, Shanghai 200438, China}
\affiliation{Hefei National Laboratory, Hefei, Anhui 230088, China}

\date{\today}

\noindent {\bf Supplemental Material for}

\maketitle


\renewcommand{\thesection}{S\arabic{section}}
\renewcommand{\thesubsection}{S\arabic{section}.\arabic{subsection}}
\renewcommand{\thesubsubsection}{S\arabic{section}.\arabic{subsection}.\arabic{subsubsection}}
\renewcommand{\theequation}{S\arabic{equation}}
\renewcommand{\thetable}{S\arabic{table}}
\renewcommand{\thefigure}{S\arabic{figure}}
\renewcommand{\thepage}{S\arabic{page}}

\tableofcontents
\clearpage

\section{Open Quantum System Models} \label{sec:oqs}

In this section, we outline the general framework for an open quantum system coupled to an environment of non-interacting bosonic (fermionic) states. The total Hamiltonian is
\begin{equation}
H_{\rm T}=H_{\rm S}+H_{\rm E}+H_{\rm I},
\end{equation}
with the environment initially in thermal equilibrium, \(\rho_{\rm E}^{\rm eq}=e^{-\beta H_{\rm E}}/\operatorname{tr}(e^{-\beta H_{\rm E}})\), where \(\beta=1/T\) is the inverse temperature. We set \(\hbar=k_{\rm B}=1\) throughout. The two cases, concerning bosonic and fermionic environments, are presented separately below.

\subsection{Bosonic Environment}

For an open quantum system coupled with a bosonic environment, the environment (bath) Hamiltonian and the system-bath interaction Hamiltonian are
\begin{equation}
H_{\rm E}=\sum_k \omega_k \hat{b}_k^\dagger \hat{b}_k,\qquad
H_{\rm I}=\hat{Q}\sum_k g_k(\hat{b}_k+\hat{b}_k^\dagger),
\end{equation}
where \(\hat{b}_k^\dagger\) (\(\hat{b}_k\)) creates (annihilates) a bosonic bath mode of frequency \(\omega_k\), \(\hat{Q}\) is a system operator, and \(g_k\) is the coupling strength.

The influence of the bath environment on the reduced system dynamics is fully encoded in the bath correlation function
\begin{equation}
C(t)\equiv {\rm tr}\left[\hat{F}(t)\hat{F}(0)\rho_{\rm E}^{\rm eq}\right],
\end{equation}
where \(\hat{F}(t)=e^{iH_{\rm E}t}\hat{F}e^{-iH_{\rm E}t}\) with \(\hat{F}\equiv\sum_k g_k(\hat{b}_k+\hat{b}_k^\dagger)\). Following the fluctuation-dissipation theorem, the bath hybridization correlation function is expressed as
\begin{equation}
C(t)=\frac{1}{\pi}\int_{-\infty}^{\infty}J(\omega)f_\beta(\omega) e^{-i\omega t}\,d\omega.
\end{equation}
Here, \(J(\omega)\equiv \pi\sum_k |g_k|^2\delta(\omega-\omega_k)\) defines the bath spectral density for $\omega >0$, which is extended to negative frequencies via the symmetry relation $J(\omega) = -J(-\omega)$ for $\omega < 0$. In the continuum limit, the discrete sum becomes an integral, rendering $J(\omega)$ as a continuous function of frequency. For a bosonic bath, $f_\beta(\omega) = [1-\exp(-\beta\omega)]^{-1}$ is the Bose-Einstein distribution function.

\subsection{Fermionic Environment}

For a fermionic environment, the bath Hamiltonian and the system-bath interaction Hamiltonian are
\begin{equation}
H_{\rm E}=\sum_{k,\alpha}\epsilon_{k\alpha}\hat{a}_{k\alpha}^\dagger\hat{a}_{k\alpha},\qquad
H_{\rm I}=\sum_{k,\alpha}g_k\hat{a}_{k\alpha}\hat{d}_{\alpha}^\dagger+\mathrm{H.c.},
\end{equation}
where \(\alpha=\uparrow,\downarrow\), \(\hat{a}_{k\alpha}^\dagger\) (\(\hat{a}_{k\alpha}\)) creates (annihilates) a fermion of spin-$\alpha$ on the $k$th bath state, and \(\hat{d}_\alpha^\dagger\) (\(\hat{d}_\alpha\)) creates (annihilates) a spin-\(\alpha\) fermion on the local orbital of the system.  
We define the system coupling operator \(\hat{F}_\alpha \equiv\sum_{k}g_k \,\hat{a}_{k\alpha}\).
For simplicity, we restrict our analysis to a spin-unpolarized system in this work and hereafter suppress the spin index $\alpha$ (denoting the operator simply as $\hat{F}$).

Analogous to the bosonic case, the influence of the environment is captured by bath correlation functions
\begin{equation} \label{eqn:def-ct-fermion}
C^{\sigma}(t)={\rm tr}\left[\hat{F}^\sigma(t)\hat{F}^{\bar{\sigma}}(0)\rho_{\rm E}^{\rm eq}\right], 
\end{equation}
where \(\sigma\in\{+,-\}\), \(\bar{\sigma}=-\sigma\), and $
\hat{F}^+(t)=e^{iH_{\rm E}t}\hat{F}^\dagger e^{-iH_{\rm E}t}$, $
\hat{F}^-(t)=e^{iH_{\rm E}t}\hat{F} e^{-iH_{\rm E}t}$. These bath correlation functions admit the spectral representation
\begin{equation}
C^\sigma(t)=\frac{1}{\pi}\int_{-\infty}^{\infty}J(\omega)f_{\beta}^\sigma(\omega)e^{i\sigma\omega t}\,d\omega,
\end{equation}
with  bath spectral density $J(\omega)\equiv \pi\sum_k |g_k|^2\delta(\omega-\epsilon_{k})$ and the Fermi–Dirac distribution function $f_{\beta}^\sigma(\omega)=\left\{1+\exp[\sigma\beta(\omega-\mu)]\right\}^{-1}$ (\(\mu\) the chemical potential, set to zero in the main text).

In the derivation that follows for fermionic baths, we focus on the \(C^+(t)\) branch of the bath correlation functions, denoting it by \(C(t)\) for notational brevity. The \(C^-(t)\) component is treated similarly and will be indicated explicitly where needed. Likewise, we denote \(f_{\beta}^{+}(\omega)\) as \(f_{\beta}(\omega)\) for the fermionic bath.

\subsection{Decomposition of Bath Correlation Function} \label{subsec:decomp-bcf}

If the spectral density can be analytically continued to the entire complex plane, the residue theorem will yield a decomposition of the bath correlation function into a sum of decaying exponentials. The decay rates are governed by the poles of \(J(\omega)\) and the Matsubara poles of the distribution function (all assumed distinct and simple).

For a bosonic environment, closing the contour in the lower half-plane for \(t>0\) gives
\begin{equation} \label{eqn:ct-bosonic}
C(t)=-2i\sum_{\omega_0}\mathrm{Res}[J(\omega_0)] f_{\beta}(\omega_0)e^{-i\omega_0 t}
-2i\sum_k \frac{1}{\beta}J(-i\nu_k^{\rm B})e^{-\nu_k^{\rm B} t},
\end{equation}
where $\omega_0$ are the poles of $J(\omega)$ in the lower half-plane ($\mathrm{Im}\,(\omega_0)<0$), $\mathrm{Res}[J(\omega_0)]\equiv\lim_{\omega\to\omega_0}(\omega-\omega_0)J(\omega)$ is the residue at $\omega_0$, and $\nu_k^{\rm B}=2\pi k/\beta$ ($k\in\mathbb{Z}^+$) are the bosonic Matsubara frequencies with $\mathrm{Res}[f_{\beta}(-i\nu_k^{\rm B})]=1/\beta$. 

For a fermionic environment, closing the contour in the upper half-plane gives
\begin{equation} \label{eqn:ct-fermionic}
C(t)=2i\sum_{\omega_0}\mathrm{Res}[J(\omega_0)] f_{\beta}(\omega_0)e^{i\omega_0 t}-2i\sum_k \frac{1}{\beta}J(i\nu_k^{\rm F})e^{-\nu_k^{\rm F} t},
\end{equation}
where \(\omega_0\) are the poles of \(J(\omega)\) in the upper half-plane ($\mathrm{Im}\,(\omega_0)>0$), and \(\nu_k^{\rm F}=(2k+1)\pi/\beta\) are the fermionic Matsubara frequencies with \(\mathrm{Res}[f_{\beta}(i\nu_k^{\rm F})]=-1/\beta\).

At low temperatures, when the lowest Matsubara frequency is much smaller than the imaginary parts of the poles of \(J(\omega)\), i.e., $2\pi/\beta \ll \min_{\omega_0} |\mathrm{Im}(\omega_0)|$, the Matsubara contribution yields the slowest-decaying component of the bath correlation function. The long-time asymptotic form is then $C(t)\sim e^{-t/\tau_c}$ with $\tau_c\propto \beta$, where $\sim$ denotes the long-time asymptotic behavior with prefactors omitted.

In the zero-temperature limit, the Matsubara sums in Eqs.~\eqref{eqn:ct-bosonic} and \eqref{eqn:ct-fermionic} become an integral and the bath correlation functions exhibit algebraic decay. 

For a bosonic environment,
\begin{equation}
C(t)=-2i\sum_{\omega_0}\mathrm{Res}[J(\omega_0)] f_{\beta}(\omega_0)e^{-i\omega_0 t}-\frac{i}{\pi}\int_0^\infty J(-ix)e^{-xt}\,dx,
\end{equation}
which in the long-time limit behaves as
\begin{equation}
C(t)\sim \frac{1}{t^{\alpha+1}}.
\end{equation}
Here, $J(-ix)\propto x^{\alpha}$ in the low-energy regime $x\to0^+$  with $\alpha\ge0$~\cite{Giraldi2017,Aha18104306,Bha22128226}. 

For a fermionic environment,
\begin{equation}
C(t)=2i\sum_{\omega_0}\mathrm{Res}[J(\omega_0)] f_{\beta}(\omega_0)e^{i\omega_0 t}-\frac{i}{\pi}\int_0^\infty J(ix)e^{-xt}\,dx,
\end{equation}
and correspondingly,
\begin{equation}
C(t)\sim \frac{1}{t}.
\end{equation}

\section{Strategy for Enhancing non-Markovianity}

The non-Markovianity of the reduced dynamics can be enhanced by truncating the spectral density. In this section, we introduce the truncated bath spectral density function and its softened version, and derive the corresponding bath correlation functions. 

\subsection{Truncating Spectral Density}

In Section~\ref{sec:oqs}, the slowest-decaying correlation is obtained only in the zero-temperature limit. To achieve similar behavior at finite temperature, we introduce a sharp cutoff in the spectral density:
\begin{equation}
    J_{\rm trun}(\omega)=\Theta(|\omega|-W_0)J_0(\omega),
\end{equation}
where $\Theta(\omega\geq0)=1$, $\Theta(\omega<0)=0$ and $W_0$ denotes the truncation width.

\subsubsection{Bosonic Bath}
For a bosonic bath, the bath correlation function corresponding to the truncated spectral density is expressed as
\begin{align}
C_{\rm trun}(t)=&\frac{1}{\pi}\int_{-\infty}^\infty \frac{\Theta(|\omega|-W_0)J_0(\omega)}{1-\exp(-\beta\omega)}e^{-i\omega t}\,d\omega   \nonumber \\ 
=&\frac{1}{\pi} \lim_{\beta'\rightarrow\infty}\int_{-\infty}^\infty \left[\frac{1}{1+\exp[\beta'(\omega+W_0)]}+\frac{1}{1+\exp[\beta'(-\omega+W_0)]}\right]\frac{J_0(\omega)e^{-i\omega t}}{1-\exp(-\beta\omega)}\,d\omega   \nonumber \\ 
=&-2i\sum_{\omega_0}\frac{\Theta(|\omega_0|-W_0)}{1-\exp(-\beta\omega_0)}{\rm Res}[J_0(\omega_0)]e^{-i\omega_0 t}   \nonumber \\ 
&-2i\lim_{\beta'\rightarrow\infty} \sum_k\frac{1}{\beta'}\left[\frac{J_0(W_0-ix_k)e^{-iW_0t-x_k t}}{1-\exp[-\beta(W_0-ix_k)]}-\frac{J_0(-W_0-ix_k)e^{iW_0t-x_kt}}{1-\exp[-\beta(-W_0-ix_k)]}\right]   \nonumber \\ 
&-2i\lim_{\beta'\rightarrow\infty}\sum_k\frac{J_0(-i\nu_k^{\rm B})e^{-\nu_k^{\rm B} t}}{\beta} \left[\frac{1}{1+\exp[\beta'(-i\nu_k^{\rm B}+W_0)]}+\frac{1}{1+\exp[\beta'(i\nu_k^{\rm B}+W_0)]}\right]   \nonumber \\ 
=&-2i\sum_{\omega_0}\frac{\Theta(|\omega_0|-W_0)}{1-\exp(-\beta\omega_0)}{\rm Res}[J_0(\omega_0)]e^{-i\omega_0 t}  \nonumber\\
&-2i\lim_{\beta'\rightarrow\infty} \sum_k\frac{1}{\beta'}\left[\frac{J_0(W_0-ix_k)e^{-iW_0t-x_k t}}{1-\exp[-\beta(W_0-ix_k)]}-\frac{J_0(-W_0-ix_k)e^{iW_0t-x_kt}}{1-\exp[-\beta(-W_0-ix_k)]}\right]  \nonumber\\
=&-2i\sum_{\omega_0}\frac{\Theta(|\omega_0|-W_0)}{1-\exp(-\beta\omega_0)}{\rm Res}[J_0(\omega_0)]e^{-i\omega_0 t} \nonumber \\
&-\frac{i}{\pi}\int_0^\infty \left[\frac{J_0(W_0-ix)e^{-iW_0t-x t}}{1-\exp[-\beta(W_0-ix)]}-\frac{J_0(-W_0-ix)e^{iW_0t-xt}}{1-\exp[-\beta(-W_0-ix)]}\right]dx,
\label{eq:truncatedC}
\end{align}
where \(\omega_0\) are the poles of \(J_0(\omega)\) in the lower half-plane, as defined previously. In the second equality of Eq.~(\ref{eq:truncatedC}), we have used the representation
\begin{equation}
\Theta(|\omega|-W_0)=\lim_{\beta'\rightarrow\infty}\left\{ \frac{1}{1+e^{\beta'(\omega+W_0)}}+\frac{1}{1+e^{\beta'(-\omega+W_0)}}\right\},
\end{equation}
where the convergence is pointwise (except at the discontinuities $\omega=\pm W_0$). Here \(\beta'\) is an auxiliary parameter used only for contour integration and unrelated to the physical temperature \(\beta\); it induces poles at $x_k=i(2k+1)\pi/\beta'$. Importantly, the last term on the right-hand side of the third equality of Eq.~(\ref{eq:truncatedC}) becomes negligible because
\begin{equation}
\lim_{\beta'\rightarrow \infty}\frac{1}{1+\exp[\beta'(i\nu_k^{\rm B}+W_0)]}=\lim_{\beta'\rightarrow \infty}\frac{1}{1+\exp[\beta'(-i\nu_k^{\rm B}+W_0)]} = 0
\end{equation}
for \(W_0>0\).

The step function $\Theta(\vert\omega\vert-W_0)$ introduces new poles appear along the lines $\omega = \pm W_0 - ix$ ($x>0$) in the lower half-plane, corresponding to the truncation edges.  These edge poles give the explicit asymptotic form of the bath correlation function in the long-time limit:
\begin{equation}
C_{\rm trun}(t) \sim \left(\eta_-e^{-iW_0 t}+\eta_+e^{iW_0 t}\right)\frac{1}{t}.   \label{eqn:ctrun-boson}
\end{equation}
Here, $\eta_-=-iJ_0(W_0)/[\pi(1-e^{-\beta W_0})]$ and $\eta_+=iJ_0(-W_0)/[\pi(1-e^{\beta W_0})]$. Based on the symmetry relation of the bosonic spectral density function, $J_0(W_0)=-J_0(-W_0)$, one finds $\eta_-+\eta_+=-iJ_0(W_0)/\pi$.
Clearly, Eq.~\eqref{eqn:ctrun-boson} recovers Eq.~(4) in the main text under the condition $T \ll W_0$.

\subsubsection{Fermionic Bath}

For a fermionic bath, the bath correlation function corresponding to a truncated bath spectral density function is expressed as
\begin{align}
C_{\rm trun}(t)=&\frac{1}{\pi}\int_{-\infty}^\infty \frac{\Theta(|\omega|-W_0)J_0(\omega)}{1+\exp(\beta\omega)}e^{i\omega t}\,d\omega \nonumber\\
=&\frac{1}{\pi} \lim_{\beta'\rightarrow\infty}\int_{-\infty}^\infty \left[\frac{1}{1+\exp[\beta'(\omega+W_0)]}+\frac{1}{1+\exp[\beta'(-\omega+W_0)]}\right]\frac{J_0(\omega)e^{i\omega t}}{1+\exp(\beta\omega)}\,d\omega  \nonumber\\
=&2i\sum_{\omega_0}\frac{\Theta(|\omega_0|-W_0)}{1+\exp(\beta\omega_0)}{\rm Res}[J_0(\omega_0)]e^{i\omega_0 t}  \nonumber\\
&+\frac{i}{\pi}\int_0^\infty \left[\frac{J_0(W_0+ix)e^{iW_0t-x t}}{1+\exp[\beta(W_0+ix)]}-\frac{J_0(-W_0+ix)e^{-iW_0t-xt}}{1+\exp[\beta(-W_0+ix)]}\right]dx.
\label{eq:truncatedC_F+}
\end{align}
Here, \(\omega_0\) are the poles of \(J(\omega)\) in the upper half-plane. The long-time behavior of the bath correlation function is governed by the edge contributions:
\begin{equation} \label{eqn:c-trun-fermion-1}
    C_{\rm trun}(t) \sim \left(\eta_-e^{-iW_0 t}+\eta_+e^{iW_0 t}\right)\frac{1}{t},  
\end{equation}
where $\eta_-=-iJ_0(-W_0)/[\pi(1+e^{-\beta W_0})]$ and $\eta_+=iJ_0(W_0)/[\pi(1+e^{\beta W_0})]$. If $J_0(W_0)=J_0(-W_0)$, we have $\eta_+-\eta_-=iJ_0(W_0)/\pi$. 

As introduced in Section~S1.2, the fermionic bath correlation functions have two branches, $C^+(t)$ and $C^-(t)$, defined in Eq.~\eqref{eqn:def-ct-fermion}. For notational brevity, we have set $C(t) \equiv C^+(t)$, so that $C_{\rm trun}(t)$ given by Eq.~\eqref{eqn:c-trun-fermion-1} denotes $C^+_{\rm trun}(t)$. Similarly, $C_{\rm trun}^-(t)$ is given by
\begin{align}
C_{\rm trun}^-(t)=&\frac{1}{\pi}\int_{-\infty}^\infty \frac{\Theta(|\omega|-W_0)J_0(\omega)}{1+\exp(-\beta\omega)}e^{-i\omega t}\,d\omega \nonumber\\
=&\frac{1}{\pi} \lim_{\beta'\rightarrow\infty}\int_{-\infty}^\infty \left[\frac{1}{1+\exp[\beta'(\omega+W_0)]}+\frac{1}{1+\exp[\beta'(-\omega+W_0)]}\right]\frac{J_0(\omega)e^{-i\omega t}}{1+\exp(-\beta\omega)}\,d\omega  \nonumber\\
=&-2i\sum_{\omega_0}\frac{\Theta(|\omega_0|-W_0)}{1+\exp(-\beta\omega_0)}{\rm Res}[J_0(\omega_0)]e^{-i\omega_0 t}  \nonumber\\
&-\frac{i}{\pi}\int_0^\infty \left[\frac{J_0(W_0-ix)e^{-iW_0t-x t}}{1+\exp[-\beta(W_0-ix)]}-\frac{J_0(-W_0-ix)e^{iW_0t-xt}}{1+\exp[-\beta(-W_0-ix)]}\right]dx.
\label{eq:truncatedC_F+}
\end{align}
In particular, if $J_0(W_0)=J_0(-W_0)$, $C_{\rm trun}^-(t)$ exhibits the same asymptotic form as $C_{\rm trun}^+(t) = C_{\rm trun}(t)$, given by Eq.~\eqref{eqn:c-trun-fermion-1}. Furthermore, under the condition $T \ll W_0$, Eq.~\eqref{eqn:c-trun-fermion-1} reduces to Eq.~(4) in the main text.

\subsection{Spectral Edge-softening Protocol}

To continuously tune the decay behavior of the bath correlation function, we convolve the truncated bath spectral function $J_{\rm trun}$ with a normalized smearing kernel $g(\omega)$:
\begin{equation} \label{eq:Lorentz_moll}
J(\omega) = (J_{\rm trun}*g)(\omega)=\int_{-\infty}^\infty J_{\rm trun}(\omega') g(\omega-\omega') \,d\omega'.
\end{equation}
This convolution preserves the total integral of $J(\omega)$:
\begin{equation}\label{eq:conv_conse}
    \int_{-\infty}^\infty J(\omega)\,d\omega =\int_{-\infty}^\infty J_{\rm trun}(\omega)\,d\omega.
\end{equation}

We now ask whether the same conservation holds for the total hybridization strength, $\frac{1}{\pi}\int J(\omega) f_\beta(\omega)\,d\omega$, when $J$ is weighted by distribution function $f_\beta$. By choosing a sufficiently narrow kernel relative to the gap in $J_{\rm trun}$, the gap width of the softened $J$ is almost unchanged. This makes, under the low temperature $\exp(W_0/T)\gg1$, the distribution functions effectively become step functions when multiplied by the bath spectral density: $J(\omega)f_\beta(\omega)=J(\omega)\Theta(\omega)$ for bosons, and $J(\omega)f_\beta(\omega)=J(\omega)[1-\Theta(\omega)]$ for fermions ($f_\beta$ taking $f_\beta^+)$. 
Furthermore, the narrow kernel ensures the convolution in Eq.~\eqref{eq:Lorentz_moll} does not mix positive and negative frequency regions, allowing step functions to commute with the convolution:
\begin{equation}\label{eq:half_conv}
(J_{\rm trun}\Theta) * g = (J_{\rm trun} * g)\Theta,\qquad[J_{\rm trun}(1-\Theta)] * g = (J_{\rm trun} * g)(1-\Theta).
\end{equation}
Applying this to $J(\omega)f_\beta(\omega)$ gives
\begin{equation}\label{eq:commu}
 (J_{\rm trun}f_\beta) * g = (J_{\rm trun} * g)f_\beta\;.   
\end{equation}
Thus, convolving $J_{\rm trun}$ is equivalent to convolving $J_{\rm trun}f_\beta$ for the correlation function.  Integrating Eq.~\eqref{eq:commu} over frequency and using Eq.~\eqref{eq:conv_conse}, we arrive at
\begin{equation}
    \int_{-\infty}^{\infty} \frac{J(\omega)}{\pi}f_{\beta}(\omega)\,d\omega
    =
    \int_{-\infty}^{\infty} \frac{J_{\rm trun}(\omega)}{\pi}f_\beta(\omega)\,d\omega.
\end{equation}
Therefore, under the conditions stated above, the total hybridization strength is conserved under the spectral edge-softening protocol. This conservation, together with the convolution theorem, further implies that the bath correlation function factorizes as
\begin{equation}
    C(t) = C_{\rm trun}(t) \, G(t),
\end{equation}
where $G(t)$ is the Fourier transform of the kernel $g(\omega)$.

For the numerical calculations, we choose a Lorentzian kernel $g(\omega)=\epsilon/[\pi(\omega^2+\epsilon^2)]$ and set $T/\omega_s=0.001$ and $W_0=\omega_s$ (where $\omega_s$ is the characteristic frequency of the system),  so that $G(t)=e^{-\epsilon t}$. The bath correlation function then decays as
\begin{equation}
    C(t)=C_{\rm trun}(t)e^{-\epsilon t}\sim \frac{e^{-iW_0 t}e^{-\epsilon t}}{t}.
\end{equation}
The coefficient $\eta_+$ of the $e^{iW_0t}$ term is heavily suppressed because the denominator of its expression contains the factor $\exp(\beta W_0)=e^{1000}$.

\begin{figure}[t]
    \centering
    \includegraphics[width=1\linewidth]{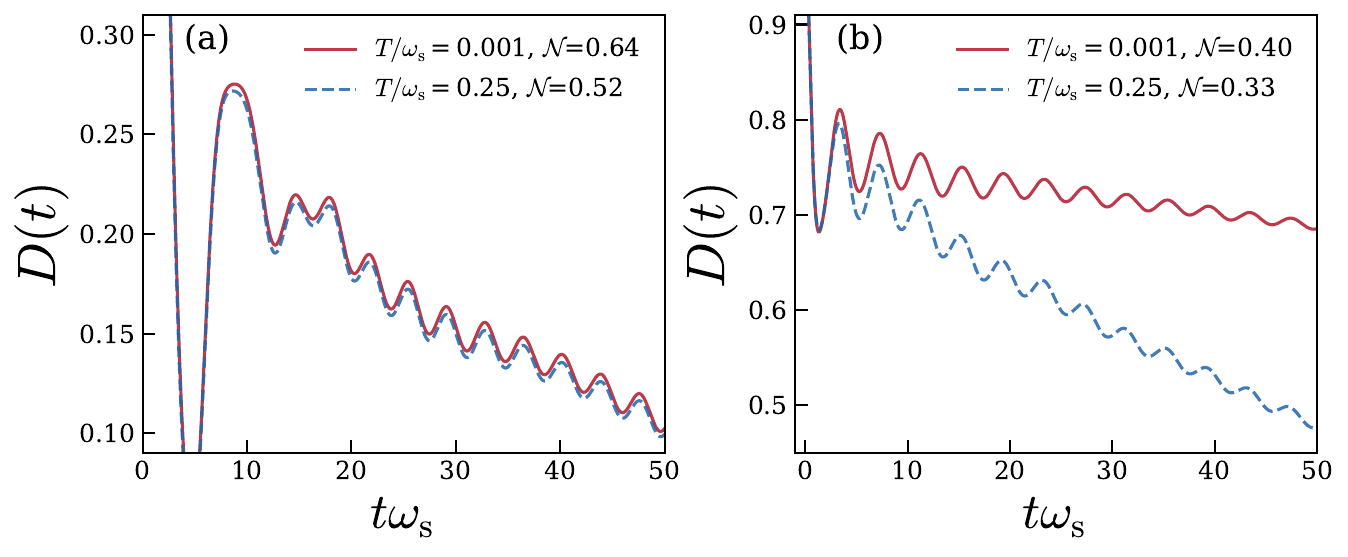}
    \caption{Trace distance dynamics at $T/\omega_s=0.001$ and $0.25$ for (a) the spin-boson model and (b) the single-impurity Anderson model, with bath spectral densities truncated at $\pm W_0$ ($W_0=\omega_s$) and softened at $\epsilon=0.0037\,\omega_s$. Non-Markovianity measures $\mathcal{N}$ are given in the legend.}
    \label{fig:Dtemp}
\end{figure}

\subsection{Temperature Dependence}

The spectral truncation can still generate sizable non-Markovianity even at relatively high temperatures, as demonstrated in Fig.~\ref{fig:Dtemp}.

In the numerical simulations, we compute the trace distance dynamics for two different temperatures in both the spin-boson model and the single-impurity Anderson model. The model parameters are the same as in Fig.~2 of the main text (with \(W_0=\omega_s\)), except that here we set the softening width to \(\epsilon=0.0037\,\omega_s\). At the low temperature $T/\omega_s=0.001$, the non-Markovianity measure is $\mathcal{N}\approx 0.64$ for the spin-boson model and $\mathcal{N}\approx 0.40$ for the Anderson model. Increasing the temperature to $T/\omega_s=0.25$ reduces it to $\mathcal{N}\approx 0.52$ and $\mathcal{N}\approx 0.33$, respectively. This indicates that, for temperatures well below the truncation width $T\ll W_0$, the non-Markovianity is insensitive to the specific value of $T$.

\section{Supplementary Figures}

\begin{figure}[t]
    \centering
    \includegraphics[width=1\linewidth]{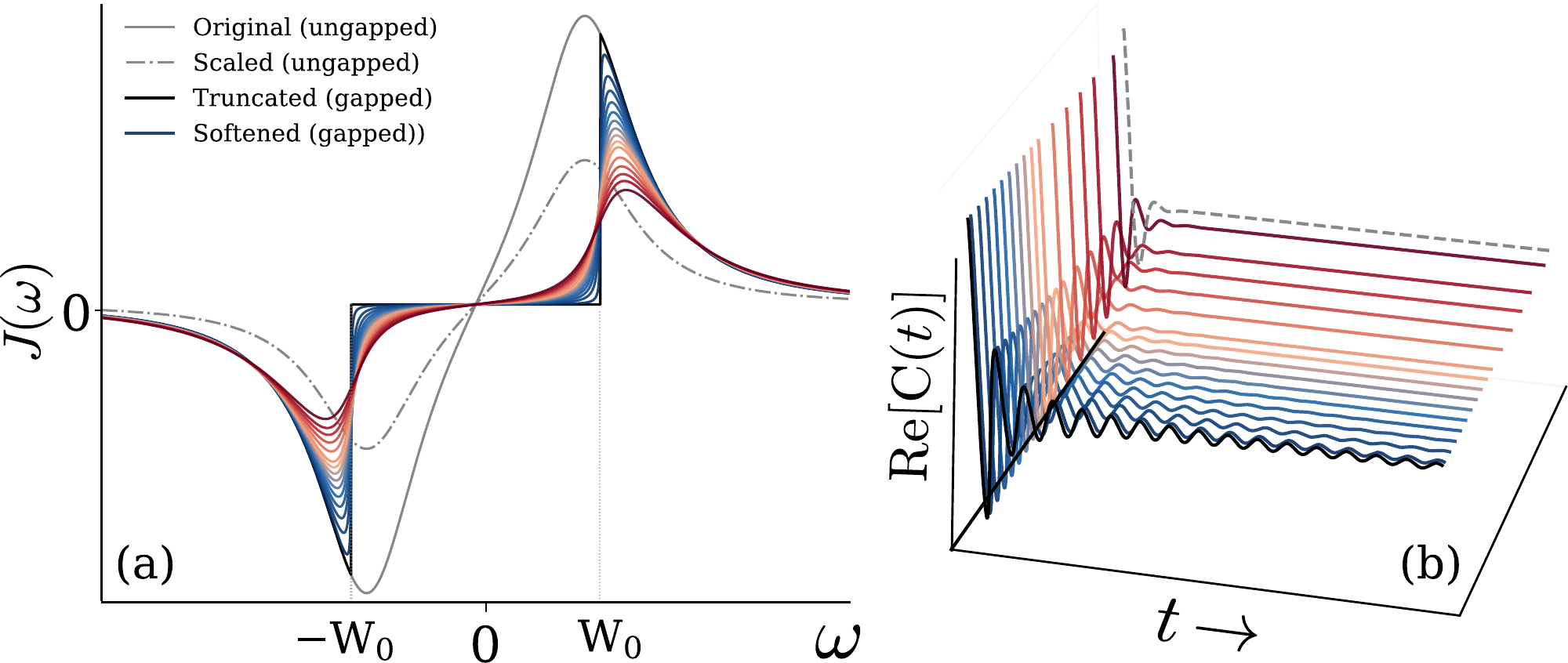}
    \caption{ Spectrum truncation and correlation functions for a bosonic bath. (a) Original spectral function (gray), truncated spectral function with a gap of width $2W_0$ (black), softened gapped spectra (colored, from blue to red with increasing softening), and a scaled ungapped spectrum (gray dash-dot). (b) Real part of the corresponding correlation functions.}
    \label{fig:schematic_B}
\end{figure}

For completeness, Fig.~\ref{fig:schematic_B} shows the bosonic counterpart of the spectral truncation and softening schematic in Fig.~1 of the main text.

\bibliography{bib}
\bibliographystyle{aip}